\documentclass[a4paper, 11pt]{article}
\usepackage{graphicx} 
\usepackage[utf8]{inputenc}
\usepackage[english]{babel}
\usepackage[T1]{fontenc}

\usepackage[a4paper, top=3cm, bottom=3cm, left=2.5cm, right=2.5cm, marginparwidth=2cm]{geometry}

\usepackage{amsmath}
\usepackage{amsfonts}
\usepackage{graphicx}
\usepackage{caption}
\usepackage{subcaption}
\usepackage[colorlinks=true, allcolors=blue]{hyperref}
\usepackage{setspace}
\usepackage{float}

\usepackage{makecell}
\usepackage[table, dvipsnames]{xcolor} 
\usepackage{multirow} 

\usepackage{doi, natbib, har2nat}
\providecommand{\keywords}[1]
{
	\small	
	\textbf{\textit{Keywords:}} #1
}

\title{Digital Asset Data Lakehouse. The concept based on a blockchain research center}
\author{Raul Cristian Bag}
\date{March 2024}

\begin{document}

\maketitle

\begin{abstract}
   In the rapidly evolving landscape of digital assets and blockchain technologies, the necessity for robust, scalable, and secure data management platforms has never been more critical. This paper introduces a novel software architecture designed to meet these demands by leveraging the inherent strengths of cloud-native technologies and modular micro-service based architectures, to facilitate efficient data management, storage and access, across different stakeholders. We detail the architectural design, including its components and interactions, and discuss how it addresses common challenges in managing blockchain data and digital assets, such as scalability, data siloing, and security vulnerabilities. We demonstrate the capabilities of the platform by  employing it into multiple real-life scenarios, namely providing data in near real-time to scientists in help with their research. Our results indicate that the proposed architecture not only enhances the efficiency and scalability of distributed data management but also opens new avenues for innovation in the research reproducibility area. This work lays the groundwork for future research and development in machine learning operations systems, offering a scalable and secure framework for the burgeoning digital economy.
\end{abstract}

\clearpage

\section{Introduction}

In the evolving landscape of digital transformation, the big volume of data generated by various digital platforms presents both an opportunity and a challenge. The vast arrays of structured and unstructured data, characterized by their significant volume, variety, and velocity—collectively known as big data—underscore the necessity for robust, scalable, and efficient data management and analysis frameworks. This paper introduces a novel approach to navigating the complexities of big data storage through the lens of blockchain and digital assets, proposing a cutting-edge software architecture tailored for the field of data lakehouses.

The explosive growth in the volume of data in the recent years, not only highlights the critical role of data in informing decision-making processes but also accentuates the challenges inherent in managing, processing, and extracting actionable insights from such vast datasets. Traditional data management systems, including relational databases and conventional data warehouses, often grapple with the scalability and agility required to accommodate the dynamic nature of big data.

Enter the concept of the data lakehouse, a paradigm that amalgamates the flexibility and scalability of data lakes with the structured and curated environment of data warehouses. The data lakehouse architecture seeks to bridge the gap between the unstructured repository of a data lake, capable of storing data in its native format, and the structured, query-optimized environment of a data warehouse. This hybrid model promises to streamline the Extract, Transform, Load (ETL) processes, thereby enhancing the efficiency of data analytics workflows and reducing the time-to-insight for businesses and researchers alike.

This paper proposes a novel software architecture designed to leverage the strengths of the data lakehouse model. We aim to address the pressing challenges of data management in the digital age, with the use-case of high-frequency trading data provided by multiple trading platforms and blockchains. Our architecture is designed to facilitate seamless data ingestion from diverse sources, including digital assets and transactions recorded on blockchain networks (e.g. binance, bitfinex, deribit, etc.), while ensuring the scalability and performance required for advanced analytics and machine learning applications.

By providing a detailed overview of the architecture's components, including data ingestion modules, blockchain connectors, and analytics engines, we aim to outline a comprehensive roadmap for researchers and practitioners looking to harness the potential of big data in the blockchain and digital assets domain.

In conclusion, the intersection of blockchain technology and data lakehouse architectures represents a frontier in the quest for efficient, secure, and scalable data management solutions. This paper not only introduces a pioneering approach to this challenge but also sets the stage for future research and development in the field. As we navigate the complexities of the digital age, the proposed architecture offers a beacon for organizations seeking to leverage big data and blockchain technology for competitive advantage and informed decision-making.
\section{Literature Review}

As already argued by \cite{chong2015big}  big data analytics plays a very important role in modern, data-driven research. However, challenges like data management, integration, and processing can pose issues that are rarely acknowledged as scientific output.

The Extract, Transform, Load (ETL) process represents a cornerstone operation within the realms of data warehousing and the facilitation of informed decision-making mechanisms.\cite{diouf2018} explains that the conventional paradigms governing the ETL process are characterized by a substantial consumption of resources, encompassing both the human capital involved in overseeing and executing these operations and the information technology resources required to sustain the computational and storage demands inherent in processing and accommodating large volumes of data. This observation underscores the need for a critical evaluation of ETL methodologies, with a view towards optimizing efficiency and minimizing the resource expenditure associated with these critical data management activities.

In the era marked by the emergence of big data, distinguished by its formidable volume, extensive variety, and accelerated velocity, the imperative for expedited processing capabilities has become progressively pronounced. The advent of cloud computing presents a potential panacea to the multifaceted challenges engendered by these characteristics of big data. Nonetheless, the prevalent "pay-per-use" billing model inherent in cloud computing infrastructures may precipitate escalated expenditures, thereby complicating its adoption as a universally viable solution. Furthermore, the deployment of proprietary Extract, Transform, Load (ETL) technologies within such environments introduces additional complexities, particularly in the context of integrating big data ecosystems. This is attributed to the inherent heterogeneity of data formats and structural paradigms across diverse data sources, which poses significant obstacles to seamless integration and efficient processing. Consequently, there is a pressing need for the development of more adaptable and cost-efficient ETL frameworks that can accommodate the dynamic nature of big data, thereby facilitating more effective and economically viable integration strategies within cloud-based architectures. \cite{diouf2018}

In addition to the aforementioned considerations, scholars have successfully constructed data lakes with the objective of advancing various research aims within the domains of financial analysis and statistical inquiry, as demonstrated in \cite{yasmin2023}, \cite{broby2019} and \cite{lee2021}. Despite these achievements, it is pertinent to note that the academic exploration of data lakes, particularly in the context of their application and utility in enhancing research methodologies and outcomes in these fields, remains substantially underexplored.

\section{Methodology}

The architectural design of a data lake requires the integration of a minimum of three distinct technological components. The first component is the storage layer, which serves as the foundational infrastructure for data accumulation. At the same level to this is the second component, the storage format or software, which dictates the modality of data storage, ensuring both the efficiency of data retrieval and the optimization of storage space. The third critical component is the software that facilitates the integration between the data source and the data lake, acting as the cohesive agent that ensures seamless data ingestion, transformation, and subsequent availability within the data lake ecosystem. This tripartite technological framework is essential for the effective realization of a data lake, enabling the aggregation, storage, and analysis of vast datasets from diversified sources in a manner that is both scalable and accessible to end-users for analytical purposes.

A critical precondition for the execution of numerous scholarly investigations is the employment of open-source technology. This stipulation arises from the recognition that proprietary solutions often incur prohibitive costs, particularly in the context of systems designed for the prolonged collection of data. In light of this consideration, the decision to utilize an S3-compatible storage provider underpinned by Ceph for the storage layer is informed by its demonstrated efficacy within large-scale data ecosystems. \cite{ceph2017} Ceph, as an open-source storage platform, offers a compelling balance of performance and cost-efficiency, thereby aligning with the economic and technical requisites of extensive data collection and analysis endeavors. This strategic choice underscores the importance of leveraging proven, open-source technologies to facilitate robust and sustainable research infrastructure, especially in scenarios requiring the accumulation and management of voluminous datasets over extended periods.

For the purposes of long-term data preservation, the adoption of Apache Parquet has been determined as the strategic choice, predicated upon its highly efficient binary format which significantly enhances performance metrics.\cite{vohra2016} This decision is further bolstered by the inherent design of Parquet as a column-oriented storage framework, which facilitates the efficient storage of both structured and unstructured data. The columnar storage model inherent to Apache Parquet optimizes both storage efficiency and the speed of data retrieval operations, thereby offering a dual advantage in the context of extensive data management. This architectural choice reflects a deliberate alignment with technologies that support the nuanced requirements of large-scale data analysis, ensuring that the storage format not only accommodates diverse data typologies but also contributes to the overall performance and scalability of the data management infrastructure.

For the Extract, Transform, Load (ETL) processes, the deployment of Apache Spark has been strategically selected, primarily due to its superior performance characteristics \cite{sparkperf}. Additionally, Apache Spark's compatibility with existing Kubernetes infrastructure presents a significant advantage, facilitating seamless integration and operational efficiency. Moreover, Apache Spark's comprehensive analytical capabilities enable direct analytics on the amassed data, thereby enhancing the utility and flexibility of the ETL framework. This multifaceted rationale underscores the selection of Apache Spark, not only for its technical prowess in handling large-scale data processing tasks but also for its congruence with contemporary cloud-native technologies and its facilitation of advanced data analysis methodologies. This decision aligns with the overarching goal of establishing a robust, scalable, and versatile data management ecosystem that can accommodate the evolving needs of data-driven research and analysis.\cite{sparkanalytics}

The existing Kubernetes infrastructure with a configuration comprising three master nodes and five worker nodes, orchestrated via Rancher Server within an OpenStack environment. The utilization of Kubernetes is of paramount importance to our current research endeavors, offering a triad of significant advantages that are critical for addressing the challenges associated with burgeoning data requirements. Firstly, Kubernetes facilitates both node and workload autoscaling, thereby ensuring that the infrastructure dynamically adapts to increasing data demands. Secondly, it provides a crucial layer of abstraction that decouples the operational logic of the code from the underlying infrastructural implementations, thereby enhancing the flexibility and scalability of the research platform. Lastly, Kubernetes significantly streamlines the management of microservices-based software architectures, offering a plethora of tools and features designed to simplify deployment, scaling, and operational tasks. This strategic employment of Kubernetes underscores its critical role in supporting the advanced computational and data management needs of contemporary research, enabling an agile, efficient, and scalable research infrastructure.
\section{A blockchain research center}

\begin{figure}[h!]
	\centering
	\includegraphics[height=7cm, keepaspectratio]{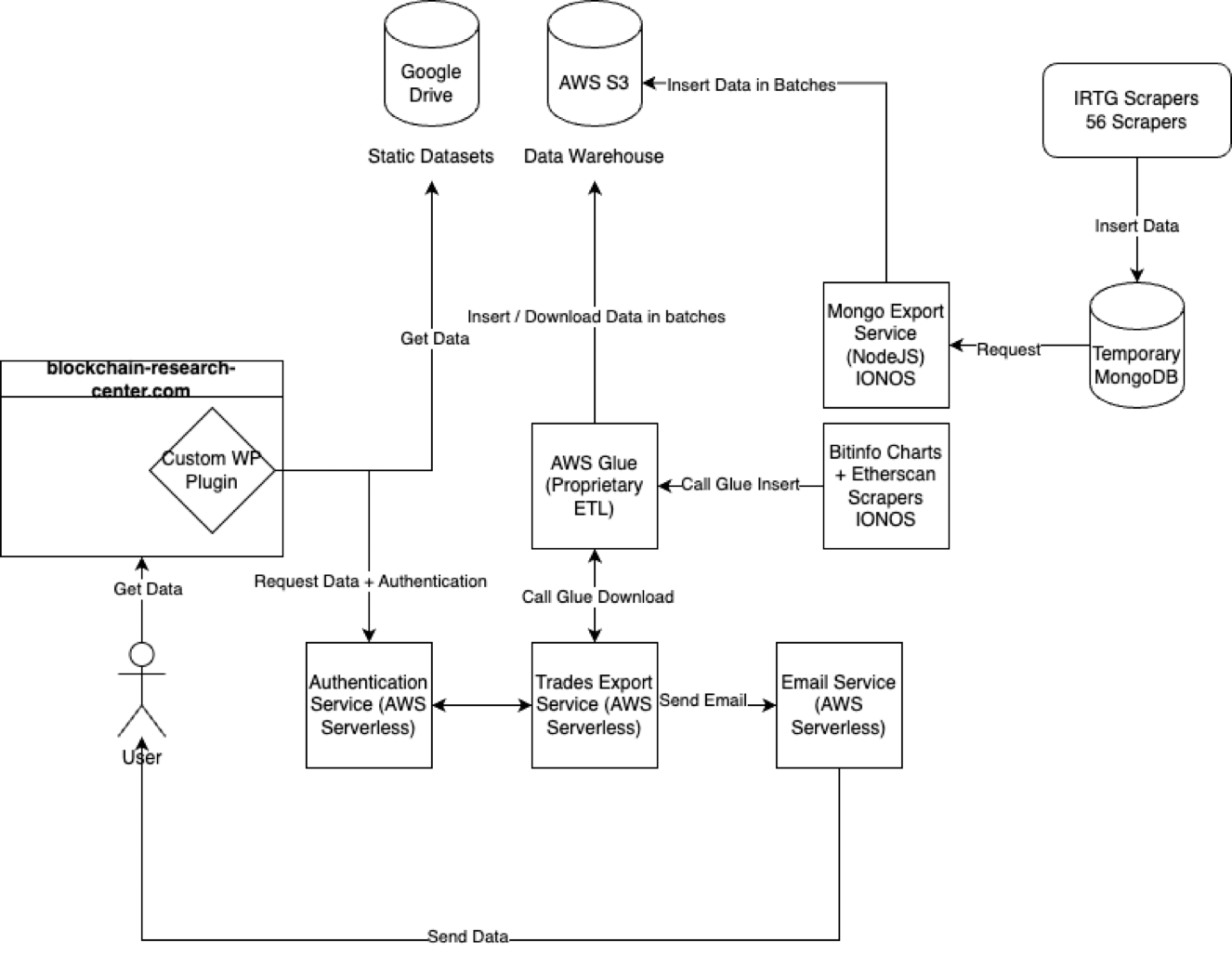}
	\caption{Previous Architecture of BRC}
	\label{fig:before_architecture}
\end{figure}

During the three years of existance to date, the \hyperlink{https://blockchain-research-center.com}{Blockchain Research Center} has helped over one thousand academics in their journey to study blockchain and other various subjects in economics and finance, by providing quality datasets that are continuously scraped, and have grown to hundreds of gigabytes along this period. It's structure before the upgrade proposed in this paper is presented in Figure \ref{fig:before_architecture}.

The diagram represents a sophisticated cloud-centric data integration architecture that orchestrates the collection, transformation, and presentation of data through a composite of cloud services and data processing workflows. Central to this system is the use of Google Drive and Amazon S3 for distributed raw data storage, alongside an extensive suite of web scrapers for data collection, emphasizing the architecture's capability to handle diverse data sources. 

Amazon Web Services (AWS) Glue is employed as a proprietary ETL service, indicating customized batch processing to accommodate the specific data transformation requirements. Data exportation is managed through a Node.js-based MongoDB export service, which interfaces with a temporary MongoDB store, representing an ephemeral staging approach for data. 

The data is ultimately consumed through a custom WordPress plugin, intended for data display on a specific domain, which, along with AWS serverless functions for authentication and email notification services, indicates a secure and scalable end-user interaction model.

The architecture depicted, while comprehensive in its integration of various data processing components, presents certain challenges that detract from its cost-effectiveness and manageability. The complexity inherent in the system arises from the amalgamation of multiple proprietary services and closed-source technologies, which may lead to increased financial overhead due to licensing fees and the specialized expertise required for operation and maintenance. 

This complexity can potentially result in diminished transparency and interoperability between components, compounding the managerial burden. The reliance on closed-source platforms restricts customization and adaptability, possibly leading to inefficient resource utilization and a concomitant escalation of operational costs. Additionally, the intricate nature of the architecture suggests a potentially steep learning curve for technical staff, further exacerbating administrative challenges and contributing to increased total cost of ownership. Therefore, while architecturally sound, the economic and administrative feasibility of the system warrants a critical evaluation, particularly in the context of long-term scalability and sustainability.

\begin{figure}[h!]
	\centering
	\includegraphics[height=7cm, keepaspectratio]{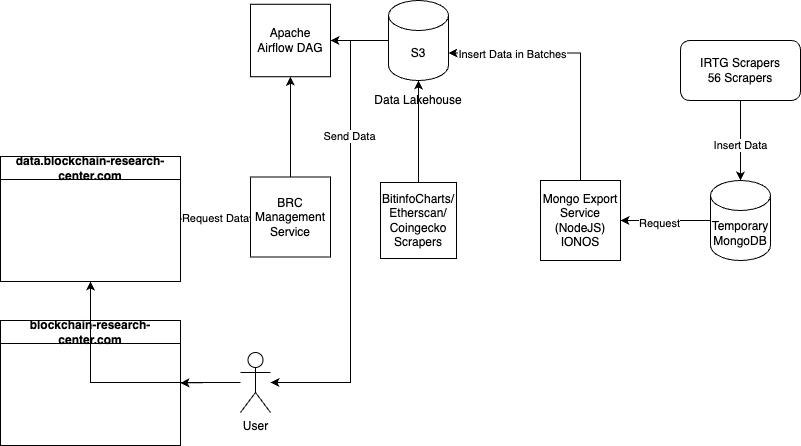}
	\caption{Proposed Architecture}
	\label{fig:after_architecture}
\end{figure}

The new architecture exhibits an uncluttered and coherent structure, ensuring a straightforward data flow from the initial acquisition by various scrapers to the final storage solutions, namely the S3 protocol with a Ceph-based implementation. This reduction in complexity is conducive to enhanced maintainability. The incorporation of well-established platforms, including Apache Airflow for workflow orchestration and Ceph Rados Gateway for S3 for data storage, contributes to the system's robustness, characterized by its resilience and capacity for maintaining data integrity.

Keeping the Mongo Export Service allows for individualized oversight and optimization of the BRC scrapers, decoupling the researchers who built them, from the system administrators. This modularity permits the isolated updating and scaling of each element, thus facilitating comprehensive system management and efficient operational monitoring.

By integrating open source technologies, the architecture avoids dependency on proprietary solutions, thereby leveraging the broad support and collective innovation of the open source community. This strategic decision may result in significant cost reductions related to licensing and enhance the system's transparency and adaptability.

Everything is hosted inside the kubernetes cluster as described in the Methdodology section, designed with horizontal scalability in mind, and is equipped to manage increasing workloads through the addition of resources or instances, particularly within its scraping and exporting components. Such a design is essential to accommodate fluctuating data processing demands.

The system's modular configuration grants a considerable degree of flexibility, enabling individual components to be modified or replaced without disrupting the overall architecture. This flexibility is instrumental in ensuring the architecture's longevity and its ability to evolve in response to emerging requirements or technological advancements.
\section{Conclusions}

In the realm of academic research, the quality of studies is intrinsically linked to the quality of the underlying data. This underscores a pivotal challenge within the academic community: the acquisition of high-quality data is often hindered by the considerable time, financial resources, and energy required, which does not directly contribute to scientific output. Additionally, the field of Data Engineering, despite its critical importance in industry settings—as significant as modeling itself—remains underappreciated in academic circles. This oversight belies the fundamental role data engineering plays in ensuring the integrity and usability of data for research purposes.

This paper has illuminated the potential of utilizing a data lakehouse architecture for managing high-frequency trading data, steering clear of blockchain for data storage to focus on the strengths of data lakehouses in handling large volumes of data efficiently. By adopting this approach, we advocate for a more refined and scalable method of data management that aligns with the rigorous demands of high-quality academic research.

The journey towards recognizing and integrating data engineering practices into academic research is crucial. Embracing data lakehouse architectures represents a forward-thinking step in addressing the contemporary challenges of data management, promising to enhance the quality of academic studies by providing a robust foundation of high-quality data. This shift not only reflects the evolving landscape of data science but also aligns academic research methodologies with industry standards, paving the way for more insightful, reliable, and impactful research outcomes.

\section*{Acknowledgement}
The language in this paper has been enhanced using Large Language Models.

\clearpage
\bibliography{bibliography}

@article{chong2015big,
  title={Big data analytics: a literature review},
  author={Chong, Dazhi and Shi, Hui},
  journal={Journal of Management Analytics},
  volume={2},
  number={3},
  pages={175--201},
  year={2015},
  publisher={Taylor \& Francis}
}

@INPROCEEDINGS{diouf2018,
  author={Diouf, Papa Senghane and Boly, Aliou and Ndiaye, Samba},
  booktitle={2018 IEEE International Conference on Innovative Research and Development (ICIRD)}, 
  title={Variety of data in the ETL processes in the cloud: State of the art}, 
  year={2018},
  volume={},
  number={},
  pages={1-5},
  doi={10.1109/ICIRD.2018.8376308}}

@InProceedings{yasmin2023,
author="Yasmin, A.
and Kamalakkannan, S.",
editor="Smys, S.
and Lafata, Pavel
and Palanisamy, Ram
and Kamel, Khaled A.",
title="Analytical Performance in Data Lake Storage of Big Data Analytics by Databricks Delta Lake for Stock Market Analysis",
booktitle="Computer Networks and Inventive Communication Technologies",
year="2023",
publisher="Springer Nature Singapore",
address="Singapore",
pages="213--226",
isbn="978-981-19-3035-5"
}

@techreport{broby2019,
title = "Creating a Financial Data Lake for Academic Fintech Research",
author = "Daniel Broby and Huckleberry Hopper",
year = "2019",
month = nov,
day = "14",
language = "English",
publisher = "University of Strathclyde",
address = "United Kingdom",
type = "WorkingPaper",
institution = "University of Strathclyde",
}

@InProceedings{lee2021,
author="Leung, Carson K.",
editor="Lee, Wookey
and Leung, Carson K.
and Nasridinov, Aziz",
title="Data Science for Big Data Applications and Services: Data Lake Management, Data Analytics and Visualization",
booktitle="Big Data Analyses, Services, and Smart Data",
year="2021",
publisher="Springer Singapore",
address="Singapore",
pages="28--44",
isbn="978-981-15-8731-3"
}

@InProceedings{ceph2017,
author="Yang, Chao-Tung
and Chen, Cai-Jin
and Chen, Tzu-Yang",
editor="Kim, Kuinam
and Joukov, Nikolai",
title="Implementation of Ceph Storage with Big Data for Performance Comparison",
booktitle="Information Science and Applications 2017",
year="2017",
publisher="Springer Singapore",
address="Singapore",
pages="625--633",
isbn="978-981-10-4154-9"
}

@inbook{vohra2016,
author="Vohra, Deepak",
title="Apache Parquet",
bookTitle="Practical Hadoop Ecosystem: A Definitive Guide to Hadoop-Related Frameworks and Tools",
year="2016",
publisher="Apress",
address="Berkeley, CA",
pages="325-335",
isbn="978-1-4842-2199-0",
doi="10.1007/978-1-4842-2199-0_8"
}

@INPROCEEDINGS{sparkperf,
  author={Wang, Kewen and Khan, Mohammad Maifi Hasan},
  booktitle={2015 IEEE 17th International Conference on High Performance Computing and Communications, 2015 IEEE 7th International Symposium on Cyberspace Safety and Security, and 2015 IEEE 12th International Conference on Embedded Software and Systems}, 
  title={Performance Prediction for Apache Spark Platform}, 
  year={2015},
  volume={},
  number={},
  pages={166-173},
  doi={10.1109/HPCC-CSS-ICESS.2015.246}}

@inproceedings{sparkanalytics,
author = {Shanahan, James G. and Dai, Laing},
title = {Large Scale Distributed Data Science Using Apache Spark},
year = {2015},
isbn = {9781450336642},
publisher = {Association for Computing Machinery},
address = {New York, NY, USA},
url = {https://doi.org/10.1145/2783258.2789993},
doi = {10.1145/2783258.2789993},
booktitle = {Proceedings of the 21th ACM SIGKDD International Conference on Knowledge Discovery and Data Mining},
pages = {2323–2324},
numpages = {2},
location = {Sydney, NSW, Australia},
series = {KDD '15}
}

\end{document}